\documentclass[%
 reprint,
 groupedaddress,
 amsmath,amssymb,
 aps,
]{revtex4-2}

\usepackage{graphicx}
\usepackage{dcolumn}
\usepackage{bm}
\usepackage{color}
\usepackage{qcircuit}
\usepackage{booktabs,eqparbox}
\usepackage[utf8]{inputenc}
\usepackage{pgfplots}
\DeclareUnicodeCharacter{2212}{−}
\usepgfplotslibrary{groupplots,dateplot}
\usetikzlibrary{patterns,shapes.arrows}
\pgfplotsset{compat=newest}

\begin{document}

\title{Variational Quantum Eigensolver Ansatz for the $J_1$-$J_2$-model}

\author{Verena Feulner}
\email{verena.vf.feulner@fau.de}
\affiliation{Physics Department, Friedrich-Alexander-Universität Erlangen Nürnberg, Germany}

\author{Michael J. Hartmann}
 \email{michael.j.hartmann@fau.de}
\affiliation{Physics Department, Friedrich-Alexander-Universität Erlangen Nürnberg, Germany}

\date{\today}

\begin{abstract}
The ground state properties of the two-dimensional
$J_1-J_2$-model are very challenging to analyze via classical numerical methods due to the high level of frustration. This makes the model a promising candidate where quantum computers could be helpful and possibly explore regimes that classical computers cannot reach.
The $J_1-J_2$-model is a quantum spin model composed of Heisenberg interactions along the rectangular lattice edges and along diagonal edges between next-nearest neighbor spins.
We propose an ansatz for the Variational Quantum Eigensolver (VQE) to approximate the ground state of an antiferromagnetic $J_1-J_2$-Hamiltonian for different lattice sizes and different ratios of $J_1$ and $J_2$.  
Moreover, we demonstrate that this ansatz can work without the need for gates along the diagonal next-nearest neighbor interactions. This simplification is of great importance for solid state based hardware with qubits on a rectangular grid, where it eliminates the need for SWAP gates.
In addition, we provide an extrapolation for the number of gates and parameters needed for larger lattice sizes, showing that these are expected to grow less than quadratically in the qubit number up to lattice sizes which  eventually can no longer be treated with classical computers.

\end{abstract}

\maketitle

\section{\label{sec:intro}Introduction}

The development of quantum hardware has made significant progress in recent years and gate sequences that are impossible or at least extremely challenging to be simulated classically \cite{48651, PhysRevLett.127.180501} have been realized. 
These gate sequences were designed for benchmark experiments and do not directly lead to ``real world" applications of interest.
Yet these achievements started the era of ``Noisy Intermediate Scale Quantum Computers (NISQ)"  which gives rise to the key question of whether useful applications of quantum computers can be possible without quantum error correction \cite{RevModPhys.94.015004}. 

A class of algorithms, that have been identified as suited for NISQ conditions, are variational quantum algorithms \cite{Cerezo_2021, VQE_1, VQE_2}. These consist of a parametrized gate sequence, for which the gate-parameters are optimized such that the energy expectation value for a considered Hamiltonian is minimized for the prepared quantum state. Two aspects make these algorithms suited for NISQ conditions. One is the fact, that rather short gate sequences can generate highly complex quantum states \cite{48651, PhysRevLett.127.180501}. The other is that the optimization uses an energy expectation value as the cost function and thus involves an average over a lot of measurements, leading to some robustness against errors.

Variational quantum algorithms have been considered for applications in quantum chemistry \cite{McArdle_2020}, where the fermionic degrees of freedom need to be mapped onto qubits via suitable transformations ensuring the anticommutation relations of fermions. Spin lattice systems in turn allow for a more direct representation on quantum computing hardware. Here, variational quantum eigensolver (VQE) algorithms have for example been considered for spin models on Kagome and square-octagon lattices  \cite{02175,08086,13375,13883} as well as one-dimensional chains \cite{symm_adaptedVQE}.

\begin{figure}[h!]
    \centering
    \includegraphics[width=0.3\textwidth]{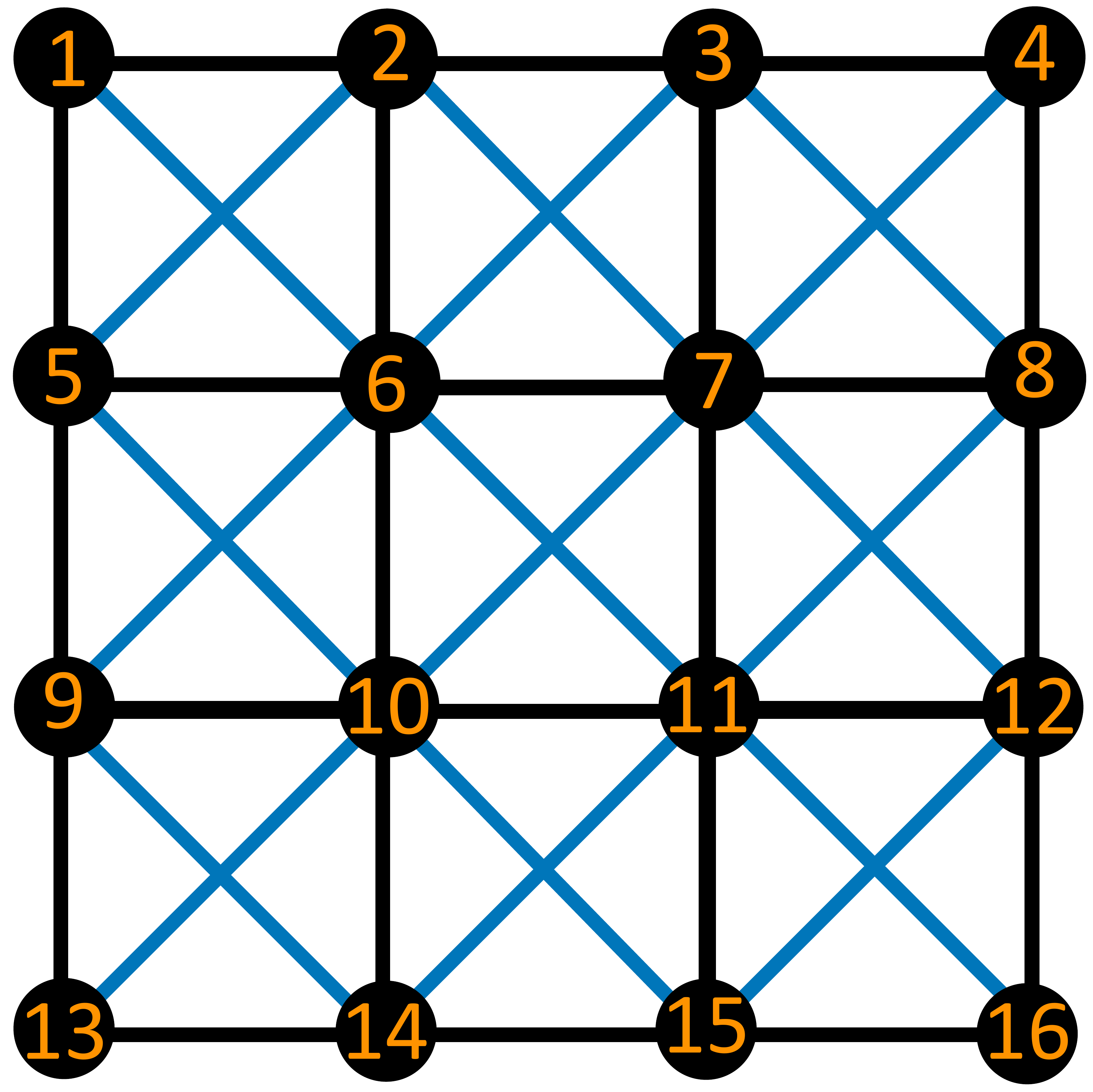}
    \caption{Interaction geometry for the $J_1-J_2$-model on a 16 qubit lattice with open boundary conditions. The blue lines visualize the next-nearest neighbor interactions with a coupling $J_2$ and the black lines the nearest neighbor interactions with a coupling $J_1$.}
    \label{fig:16grid}
\end{figure}

A model that is particularly suited for being represented on a rectangular grid of qubits, but that at the same time poses significant challenges to classical numerics, is the $J_1-J_2$-model \cite{PhysRevB.88.060402, QMC_Nakamura, PhysRevB.84.174407, Hangleiter_2020,Hasik_2021}. Indeed, several developers of superconducting qubit architectures develop rectangular grids that are forward compatible with the surface code architecture for quantum error correction. These are particularly suited for computations for spin lattice models on this type of lattices. The $J_1-J_2$-model is a spin model on a rectangular lattice that however features additional anti-ferromagnetic interactions across the diagonals of each plaquette, see figure \ref{fig:16grid}. It  can for example be used to describe CuO\textsubscript{2} planes in high-$T_c$ cuprate superconductors \cite{Mikheyenkov} or layered materials as Li\textsubscript{2}VO(Si,GE)O\textsubscript{4} \cite{VO(SiGe)} or VOMoO\textsubscript{4} \cite{VOMo}. The model however poses significant challenges to classical numerical approaches and for a specific strength of frustration,  $0.4 \gtrsim J_2/J_1 \lesssim 0.6$, its ground state remains subject of intense debate \cite{PhysRevB.100.125124}. 

In this work we develop ans\"atze for variational algorithms for the two-dimensional $J_1-J_2$-model, where we particularly focus on the classically hard parameter regime of  $0.4 \gtrsim J_2/J_1 \lesssim 0.6$. Importantly, we find that the diagonal interactions can well be captured without executing two qubit gates directly among the next-nearest neighbor qubits involved in these interactions. Two qubit gates along the edges of the rectangular grid, that can be implemented in a hardware efficient ansatz on architectures with nearest neighbor connectivity, suffice for good accuracy of the ground state approximation. Our ansatz is thus less demanding in terms of required qubit-qubit interactions than for example adiabatic algorithms for preparing desired ground states, which always require an implementation of all qubit-qubit interactions that are present in the considered model. The omission of the next nearest neighbor gates along the diagonals of the lattice leads to a significant reduction of the gate count, as these gates need to be sandwiched between two SWAP gates in standard architectures with nearest neighbor connectivity. 

Furthermore, we explore the scaling of the required numbers of quantum gates for reaching a desired accuracy in the ground state preparation with the number of spins or qubits in the model. For our ansatz without the gates for the next nearest neighbor interactions, we estimate here the promising scaling of $n^{1.2}$, where $n$ is the number of qubits. This scaling implies that $8\times8 = 64$ qubit lattices could be treated with circuits containing less than 400 two-qubit gates and less than 100 single qubit gates (We exclude single qubit Z-gates in this counting scince these can be done virually).

\section{\label{sec:j1j2model}The $J_1-J_2$-model}

The $J_1-J_2$-model is an extension of the Heisenberg model with additional Heisenberg-interactions between next-nearest neighbors \cite{J1J2_1,J1J2_2}. The model is described by the Hamiltonian
\begin{align}
\mathcal{H} &= - J_1\sum_{\langle i,j \rangle}  \vec{S}_i\cdot\vec{S}_j - J_2\sum_{\langle\langle i,j \rangle\rangle}  \vec{S}_i\cdot\vec{S}_j,
\end{align}
where $J_1$ is the strength of nearest-neighbor interactions ($\langle i,j \rangle$ indicates that the sum runs over pairs of nearest neighbors) and $J_2$ is the strength of next-nearest-neighbor interactions  ($\langle\langle i,j \rangle\rangle$ indicates that the sum runs over pairs of next-nearest neighbors). The operators $\vec{S}_j$ are vectors containing the three Pauli operators for spin-$1/2$ degrees of freedom, $\vec{S}_j = (X_j,Y_j,Z_j)^T$. Figure \ref{fig:16grid} shows the geometry of the spin lattice for 16 spins with open boundary conditions.

In this work the couplings $J_1$ and $J_2$ are chosen to be negative and thus, form antiferromagnetic interactions.

The $J_1-J_2$-model is a paradigm example for a highly frustrated system, even for the square lattice. Its geometric frustration means that its ground states are typically not formed by simple patterns \cite{geometric_frustration} like, for example, the Néel state, but rather form strongly correlated quantum states.
The frustration can be tuned by the ratio $J_2/J_1$. 
For the case of $J_2/J_1 < 0.2$, the model can be described with a spin-wave approximation. For $J_2/J_1 > 0.4$ however, this approximation breaks down \cite{zhong} and the magnetic order of the model disappears.
The following quantum phases of the system have so far been clearly identified:
For $J_2/J_1 \lesssim 0.4$,  the classical $(\pi,\pi)$ Néel behavior is observed.
For $J_2/J_1 \gtrsim 0.6$,   two collinear Néel ordered states with pitch vectors $q = (\pi,0)$ and $q = (0,\pi)$ are selected by an order by disorder mechanism. Here, order by disorder means that a soft Ising order parameter $\sigma = \hat{n}_1\cdot\hat{n}_2$ appears, where $\hat{n}_1$ and $\hat{n}_1$ denote the independent staggered magnetization of the two sublattices as written in \cite{Chandra}. The ground state energy is here independent of the angle between the staggered magnetizations \cite{dagotto}.

In the highly frustrated case  $0.4 \gtrsim J_2/J_1 \lesssim 0.6$,  quantum fluctuations destabilize the classical ordered ground state and lead to a disordered singlet ground state with a gap to the first magnetic excitation. Despite significant effort in exploring classical methods the ground state of the model at the maximally frustrated point $J_2/J_1 \sim 0.5$ and its physical properties remain the subject of intense debate \cite{PhysRevB.79.024409, PhysRevB.99.100405, PhysRevB.100.125124, Nomura_2021, Nomura_2021physrevx}. So far there have been a few conflicting proposals for the ground state candidate, for instance the plaquette valence-bond state \cite{Plaquette}, the columnar valence-bond state \cite{Columnar} or a gapless spin liquid \cite{Spin_liquid, doi:10.1126/science.235.4793.1196}. Here the ability of quantum computers to generate highly entangled states already via short gate sequences may lead to an advantage provided the experimental gate fidelities reach suitable values.

\section{\label{sec:vqe} Variational quantum algorithm}

Variational quantum algorithms \cite{VQE_1,VQE_2}
are based on the variational principle in quantum mechanics, which is used to approximate the ground state of a system. This principle reads,
\begin{align}
E_0 \leq \frac{\langle\psi|\mathcal{H}|\psi\rangle}{\langle \psi|\psi \rangle},
\end{align}
and means that the smallest energy eigenvalue of the system $E_0$ is always smaller than or equal to any expectation value of its Hamiltonian. This relation gives rise to an optimization problem, in which one seeks to minimize the expectation value of $\mathcal{H}$ for a class of states to find a good approximation to the ground state of $\mathcal{H}$.

This principle is applied in the Variational Quantum Eigensolver (VQE) \cite{VQE_1, VQE_2} to approximate the ground state energy of a given Hamiltonian. 
The resulting algorithm is a hybrid algorithm that consists of two parts: a classical parameter update and a quantum energy eigenvalue evaluation.
In the quantum part of the algorithm, the expectation value of the Hamiltonian is computed by sampling from an ansatz state $|\psi(\vec{\theta})\rangle$, which is prepared on the quantum processor via a gate sequence that depends on gate parameters $\vec{\theta}$ \cite{VQE_1, VQE_2}. 
The classical parameter update of the algorithm consists of a classical optimizer, which computes the best set of parameters $\vec{\theta}$ by calling the quantum part, to approximate the sought ground state.
Thus, the quantum part evaluates the expectation value which forms the objective function for the classical optimizer. By optimizing for better and better sets of parameters, one eventually trains the quantum computer to prepare states that are very close to the ground state of the Hamiltonian.

\subsection{\label{sec:ansatz}Choosing the ansatz}

\begin{figure*}[htb]
	 	 	\centering
	 	 	\includegraphics[width =\textwidth]{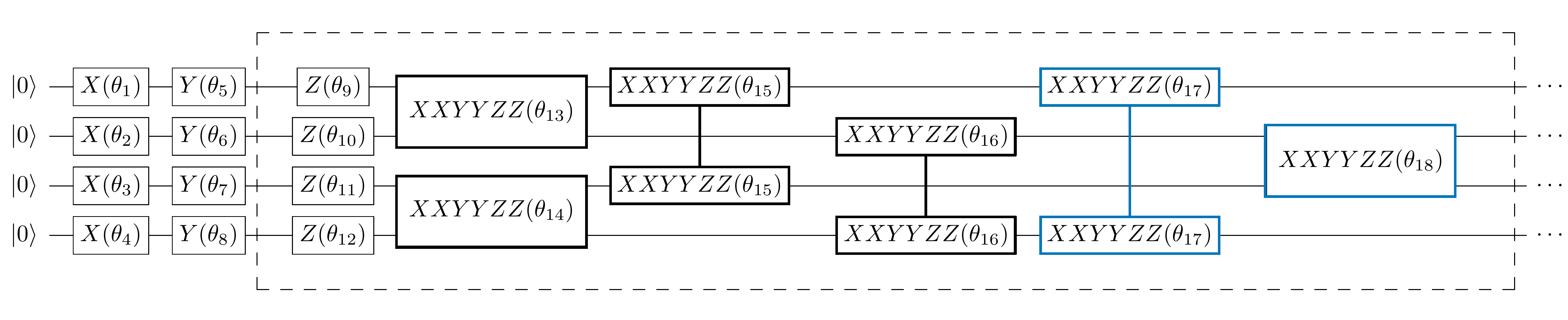}
	 	 	\caption{The ansatz used for the $J_1-J_2$-model.  Here, the first few gate-layers for a 4-qubit model are shown. The gates in the frame are repeated and we call this block a layer of gates. For the larger models, the ansatz follows the same scheme.  The first four two-qubit gates in the block (depicted by bold black frames) are the two-body gates on nearest neighbor qubits. The last two two-qubit gates in the block (depicted by bold blue frames) are the gates on next nearest neighbor qubits along the diagonal interactions. In our simulations, we found that the latter can be omitted without changing the VQE convergence.\label{fig:ansatz}}
\end{figure*}

\begin{figure}
	 	 	\centering
	 	 	\includegraphics[width =0.5\textwidth]{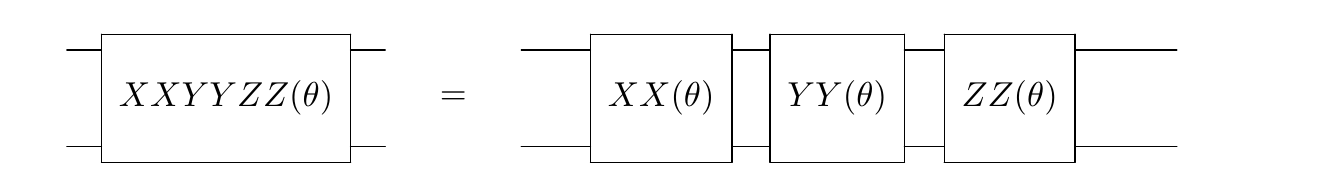}
	 	 	\caption{The parametrized ``XXYYZZ"-gate is composed of a sequence of an XX-, YY- and ZZ-gate with the same parameter.\label{gate_xxyyzz}}
\end{figure}

The ansatz we used for the $J_1-J_2$-model is sketched in figure \ref{fig:ansatz}. It consists of a parametrized X-gate and a parametrized Y-gate applied to every qubit at the beginning of the circuit. 
Afterwards, a parametrized Z- gate is applied to each qubit. 
All these single-qubit gates are parametrized by an angle $\theta$ of the rotation around the respective axis. This angle can vary from qubit to qubit. For one qubit the gates read \cite{cirq_developers}, 
 \begin{align}
X(\theta) := X^\theta & =  
   \left( \begin{array}{cc}
 G\cdot C&- i G\cdot S\\ 
 - i G\cdot S&G\cdot C  \end{array} \right)  \\
Y(\theta)  := Y^\theta & =  
   \left( \begin{array}{cc}
 G\cdot C& -G\cdot S\\ 
 G\cdot S&G\cdot C  \end{array} \right)  \\
Z(\theta) := Z^\theta & =  
   \left( \begin{array}{cc}
1&0\\ 
 0&\tilde{G} \end{array} \right) 
\end{align}
where $C=\cos(\pi \cdot\theta/2)$, $S=\sin(\pi\cdot\theta/2)$, $G =\exp(i\pi\cdot\theta/2)$ and $\tilde{G} = \exp(i\pi\cdot\theta)$.

The two-qubit gate, forming the entangling gate in the ansatz, is an ``XXYYZZ-gate'', which is applied to every edge of interactions. This ``gate'' consists of an XX-gate,  a YY-gate and a ZZ-gate, all taken to the same power $\theta$, see figure \ref{gate_xxyyzz}. These gates commute and their matrix representations read \cite{cirq_gates},
\begin{align}
  \text{XX}(\theta) &= (X \otimes X)^\theta =\left( \begin{array}{cccc}
 c& 0&0&s\\
 0&c&s&0\\
 0&s&c&0\\
 s&0&0&c
\end{array} 
 \right)\\
\text{YY}(\theta) &= (Y \otimes Y)^\theta =\left( \begin{array}{cccc}
 c& 0&0&-s\\
 0&c&s&0\\
 0&s&c&0\\
 -s&0&0&c
\end{array} 
 \right)\\
 \text{ZZ}(\theta) &= (Z \otimes Z)^\theta =\left( \begin{array}{cccc}
 1& 0&0&0\\
 0&w&0&0\\
 0&0&w&0\\
 0&0&0&1
\end{array} 
 \right) ,
\end{align}
with $ c= f\cos(\frac{\pi\cdot\theta}{2})$,  $ s= -i f\sin(\frac{\pi\cdot\theta}{2})$, $f = e^{\frac{i \pi \cdot \theta}{2}}$ and $w = e^{i\pi \cdot \theta}$.

The block formed by a layer of Z-gates and a layer of XXYYZZ-gates is then repeated until the desired convergence of the optimizer is reached.

The proposed ansatz consists of the two-qubit gates that correspond to the spin-spin interactions in the Hamiltonian \cite{symm_adaptedVQE,Liu_2019, huerga} (except for the fact that the gates on the diagonals can be left away). Using these gates has the benefit that they mutually commute. The X- and Y- single qubit gates at the beginning are chosen to mimic the unordered spin-liquid behavior. In Spin-liquids the spins are unordered due to competing interactions hence, their ground state has a high degeneracy. The spins fluctuate heavily, at low temperatures the system can ``freeze" to spin glass state \cite{PhysRevLett.114.247207, balents}. The Hamiltonian is invariant under exchange of the X, Y, and Z directions. In between the two qubit gates we however only use Z-gates since these can be implemented as virtual Z-gates \cite{virtualZZ} without cost.

Importantly, we found in our simulations that it is possible to omit the gates for the diagonal interactions with strengths $J_2$, see green lines in figure \ref{fig:16grid}. This is very useful for the implementation on superconducting quantum hardware due to the fact that superconducting qubits are only coupled to their nearest-neighbours and some superconducting circuit architectures are ordered in a rectangular grid. Thus, if one would aim for simulating the diagonal $J_2$ interactions directly via gates, SWAP-gates are needed before and after the XXYYZZ-gates. Our ansatz in turn shows that these SWAP gates can be omitted, leading to shorter circuit depth. Our approach thus also has lower hardware connectivity requirements than adiabatic ground state preparation, where all interactions in the model need to be implemented \cite{RevModPhys.90.015002}.

\subsection{Classical Optimizer \label{sec:implementation}}
For the classical optimization, the optimizer  \textit{Constrained Optimization By Linear Approximation} (COBYLA) was used with randomly chosen initial values for the variational parameters,  $-\pi \leq \theta_j \leq \pi$. This optimizing algorithm is a trust-region algorithm that aims to maintain a regular simplex during the iteration steps \cite{Powell}.
This method is nonetheless susceptible to get stuck in local minima in the energy landscape due to the difficulty of the problem.
Thus, for a few cases, if necessary, Basinhopping as implemented in SciPy was used.
This is a method that uses an arbitrary number of iterations to avoid local minima in the energy landscape, also called basins of attraction. It combines the local optimizer, with a global stepping scheme where all coordinates are displaced by a random number, called step size.  The new coordinates are accepted or rejected based on the minimal function value \cite{basinhopping}.

\section{\label{sec:results}Results}
Our main interest is to investigate achievable accuracy as well as the feasibility of our ansatz for a VQE for the simulation of spin glass models. To this end, we have simulated our VQE algorithm for lattices up to 20 spins using a classical computer. As our main interest was the suitability of the ansatz, we computed the energy expectation value of the Hamiltonian directly via the wave function and did not emulate the sampling over measured bit-strings that would be necessary when running the algorithm on a real quantum computer.

Throughout this section, if not stated otherwise, the coupling constants are fixed to the values $J_1 =-1$ and $J_2 =-0.5$. To compare the results, we investigate the achieved energy expectation value $\bar{E}$ and
\begin{align}
\frac{\bar{E}-E_0}{\text{spectral gap}}=\frac{\bar{E}-E_0}{E_1-E_0},
\label{difference_spectralgap}
\end{align}
that is the difference between the expectation value obtained from the VQE, $\bar{E}$, and the exact ground state energy $E_0$, divided by the spectral gap, which is the difference between the energy of the first excited state $E_1$ and the ground state energy $E_0$.\\

\begin{table*}
    \begin{tabular}{c|c|c|c|c}
 & number of 2 qubit gates & number of single qubit gates & number of parameters & number of layer-blocks \\
\hline
12 with diagonal gates & 609 & 108 & 312&7 \\
\hline
12 diag, period. bound. & 987 & 108 &431&7\\
\hline
12 without diag. gates & 357 &108&227&7\\
\hline
16 with diagonal gates & 861&144&432&7\\
\hline
16 without diag. gates &288&144&312&7\\
\hline
20 with diagonal gates &1944&280&931&12\\
first try smaller size& 1134&180&558&7\\
\hline
20 without diag. gates& 1080&280&640&12\\
first try smaller size& 450&140&290&5\\
\hline
    \end{tabular}
\caption{Number of gates, layers and parameters used in the different configurations of the simulation. For the variation of $J_2$ the same number of gates and layers were used as for the model with 12 qubits with diagonal interactions and without periodic boundaries. \label{tab:params}}
\end{table*}

\begin{table*}
    \centering
    \begin{tabular}{c|c|c|c|c }
         & $E_0$ & $E_1$ & $\bar{E}$ & $\frac{\bar{E}-E_0}{E_1-E_0}$ \\
\midrule
12 with diagonal gates, $J_2 =0$ &  -26.777 & -24.879 & -26.4375 & 16 \% \\
\hline
12 with diagonal gates, $J_2 =-0.5$ & -22.138 & -20.1559& -21.980& 7.9 \%\\
\hline
12 with diagonal gates, $J_2 =-2$ & -41.240 & -40.479 &  -40.8734 & 48 \% \\
\hline
12 diag, period. bound. & -25.7220 & -23.0742& -25.441& 10.6 \%\\
\hline
12 without diag. gates & -22.1380 & -20.1559& -21.995 & 7.2 \%\\
\hline
16 with diagonal gates & -30.0222& -27.8223& -29.5856& 19.9 \%\\
\hline
16 without diag. gates & -30.0222& -27.8223& -29.566 & 20.8\%\\
\hline
20 with diagonal gates &-37.7231 & -35.9921& -37.416& 17.7 \%\\
\hline
20 without diag. gates&-37.7231 & -35.9921& -37.459&  15.3 \%\\
\hline

    \end{tabular}
    \caption{The results of the various simulations. \label{tab:results}}
\end{table*}

\subsection{Various lattice sizes \label{sec:verschgroess}}
\begin{figure*}
    \centering
    \includegraphics[width=\textwidth]{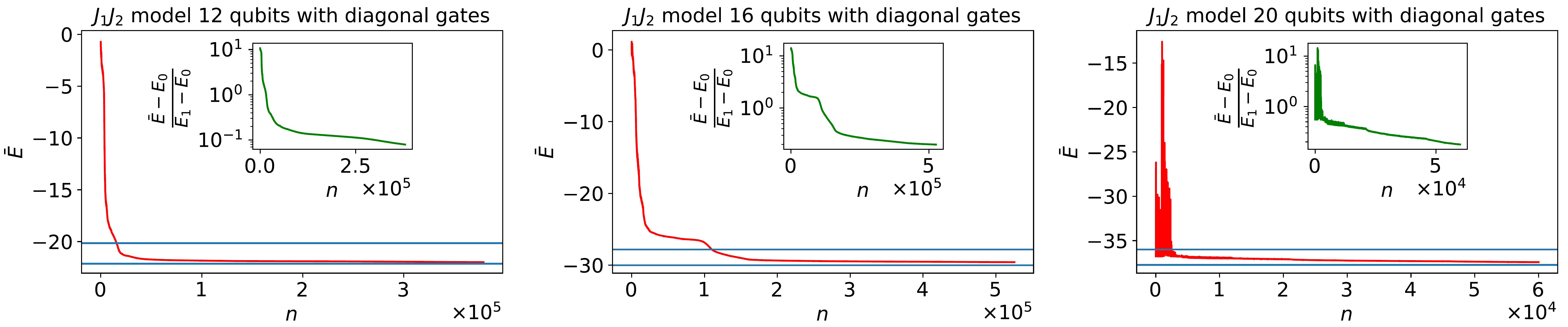}
    \caption{VQE performance for the $J_1-J_2$-model for all lattice sizes (12,16 and 20) without periodic boundary conditions and with diagonal gates. The lower blue line marks the ground state energy $E_0$ and the upper line the energy of the first excited state $E_1$. The circuit consists of the additional X- and Y-layer at the beginning plus the according number of layers, as can be seen in table \ref{tab:params}.
    The inset shows the run of the VQE-steps against the difference of the expectation value $\bar{E}$ from the VQE and the exact ground state energy $E_0$ divided by the spectral gap, which is the difference of the energy of the first excited state $E_1$ and the ground state energy.}
    \label{fig:plots_with}
\end{figure*}

\begin{figure*}
    \centering
    \includegraphics[width=\textwidth]{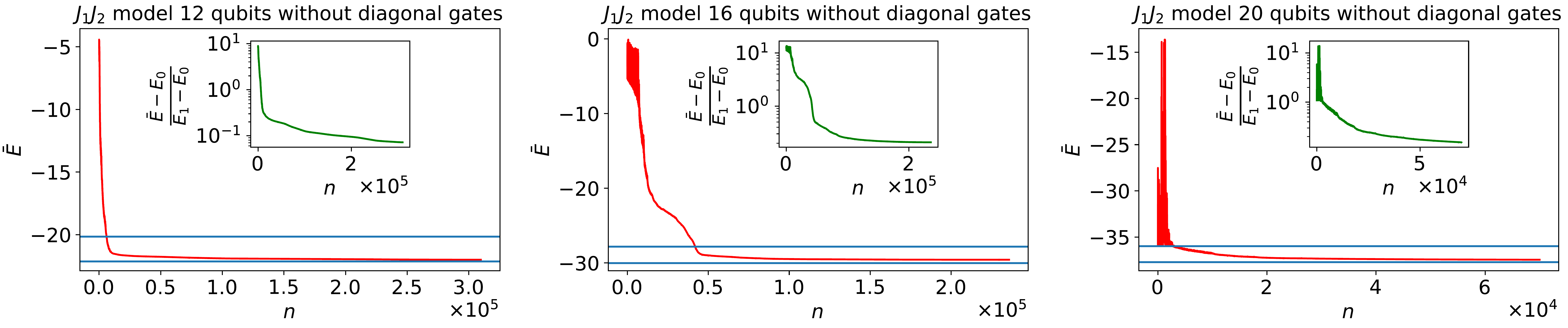}
    \caption{VQE performance for the $J_1-J_2$-model for all lattice sizes (12,16 and 20) without periodic boundary conditions and without diagonal gates. The lower blue line marks the ground state energy $E_0$ and the upper line the energy of the first excited state $E_1$. The circuit consists of the additional X- and Y-layer at the beginning plus the according number of layers, as can be seen in table \ref{tab:params}.
    The inset shows the run of the VQE-steps against the difference of the expectation value $\bar{E}$ from the VQE and the exact ground state energy $E_0$ divided by the spectral gap, which is the difference of the energy of the first excited state $E_1$ and the ground state energy.}
    \label{fig:plots_without}
\end{figure*}

We tested our ansatz with the diagonal gates and without the diagonal gates for different lattice sizes, choosing two-dimensional, rectangular lattices of 12, 16 and 20 qubits with open boundary conditions.  The results are shown in figures \ref{fig:plots_with} and \ref{fig:plots_without}. 
Table \ref{tab:params} shows the number of single and two qubit gates we used in our simulations for the numerical experiments reported in figures \ref{fig:plots_with} and \ref{fig:plots_without}.
The results $\bar{E}$ that we obtained in these numerical VQE experiments, together with the exact energies of the ground states $E_0$, the exact energies of the first excited states $E_1$ and the ratios $(\bar{E}-E_0)/(E_1-E_0)$ are reported in table \ref{tab:results}. Here, we first discuss the simulations that include the gates corresponding to the diagonal $J_2$-interactions.

\paragraph{Ansatz with diagonal gates}
For the lattice with 12 qubits with open boundary conditions, we achieve a good convergence within $10^5$ iterations. The number of layers (see figure \ref{fig:ansatz}) can be seen in table \ref{tab:params}. The VQE optimization of this lattice can be seen in figure \ref{fig:plots_with}. In this case, the ground state can be approximated with $\frac{\bar{E}-E_0}{E_1-E_0} <10 \% $.
For the 16-qubit lattice with open boundary conditions, we found the optimization result as can be seen in table \ref{tab:results} with seven gates, see table \ref{tab:params}. Thus, the VQE only ends up in the bottom  20 percent interval of the spectral gap, as can be seen in figure \ref{fig:plots_with}. To achieve better results, the ansatz can be extended to more gate layers. In the case of 16-qubits with nearest-neighbor interaction gates we also did not use the basinhopping scheme which might also help to achieve better results.

The largest lattice size we considered was a two-dimensional grid with 20 qubits. Due to the large size of the Hilbert space and the risk of running into local minima of the energy landscape, a good guess for start values of the variational parameters is beneficial. To achieve this, we first ran the VQE with seven gate layers, which resulted in a value for $\overline{E}$ that lies approximately in the middle of the spectral gap. For higher accuracy we fed the obtained set of parameters as an initial guess into our VQE ansatz with 12 ansatz layers, where we set the $\theta$-values for the additional parametrized gates to $10^{-5}$ \footnote{Setting these exactly to zero led to numerics problems, presumably because of hitting a fix-point of the optimizer.}. With this approach we get a good approximation of the ground state, albeit at the cost of doubling the number of ansatz-layers.

Overall, we find that due to the parametrized X- and Y-gates at the beginning of the ansatz, the spins in the $J_1-J_2$-model can be prepared in the expected spin liquid order for the ground state or a good approximation of it in most cases of our simulations. In all configurations, we achieve an energy result which is closer to the ground state than the first excited state. The precision of the model could be increased by more gate layers or by the use of a better suited optimizer that is less susceptible to local energy minima and the initial state.

\paragraph{Ansatz without diagonal-interaction gates 
}
To show that our ansatz works without using gates that mimic the diagonal $J_2$-interaction, we implemented the VQE without gates on the diagonals for lattices of 12, 16 and 20 qubits with open boundary conditions. The results of these simulations can also be found in table \ref{tab:results}.
The ansatz here follows the same scheme as above, except for not applying diagonal interactions via gates on next-nearest neighbor qubits (blue boxes in figure \ref{fig:ansatz}). Leaving out these gates reduces the gate count in two ways. Firstly the omitted gates need not be implemented and secondly these gates would need to be sandwiched between SWAP gates in architectures with only nearest neighbor connectivity as they cannot be implemented directly.

For this ansatz, we get a slightly better results in the optimization for all considered lattice sizes, as can be seen in figure \ref{fig:plots_without} and table \ref{tab:results}. This could be due to the fact that fewer parameters are used and the optimizer is more efficient in finding a minimum. The only case, where the value is slightly worse than for the ansatz with diagonal gates, is the 16-qubit lattice, which might be due to the fact that a better initial state for the optimization was found in the run that included gates on the diagonals.
In the cases of 12 and 16 qubits we also needed fewer iterations. In turn, the slightly increased number of iterations needed for the 20 qubit optimization is caused by the fact that only five layers of gates were here used in the ''pre-training'' stage (as compared to seven layers in the case with gates on the diagonals). Hence, the training of the full 12 layer circuit in the second training stage required more iterations.

\begin{figure*}[htb]
    \centering
    \includegraphics[width=\textwidth]{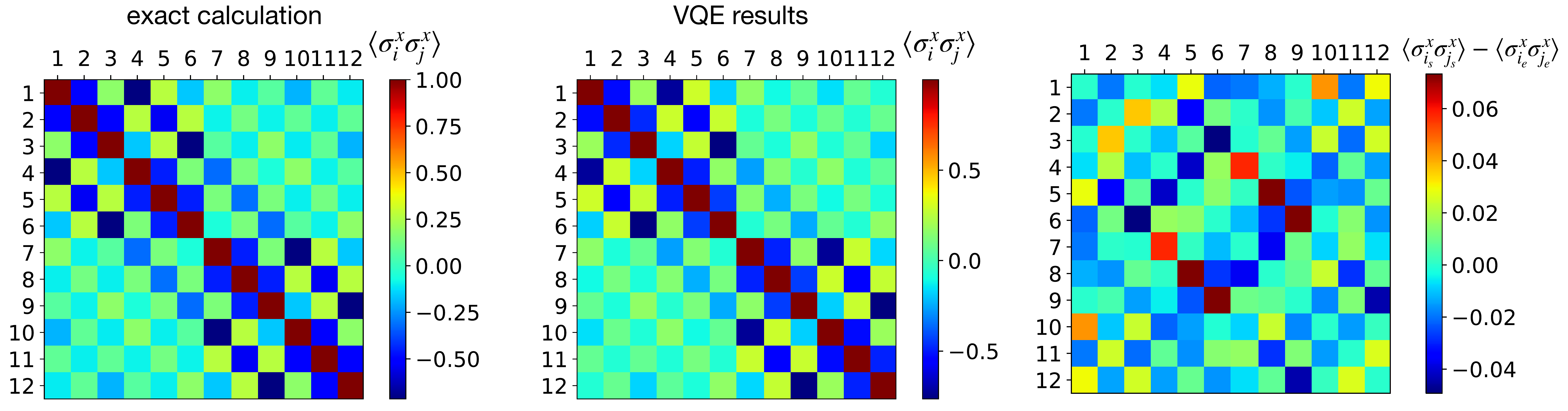}
    \caption{Spin-spin correlation in $x$-direction $\langle \sigma^x_i \sigma^x_j \rangle$ for the 12-qubit lattice without diagonal gates computed from the final state of the corresponding simulation reported in table \ref{tab:results} and figure \ref{fig:plots_without}. The plot on the left shows the correlation function calculated with the VQE-results. The middle plot is the correlation function of the exact results, and on the right the difference of both can be seen. Subscript $s$ are the results of the VQE simulation and $e$ are the exact results. }
    \label{fig: Correlation_without_NN}
\end{figure*}

\paragraph{Further ground state properties}
With the wave function output of our full wave function VQE simulation, we can also calculate expectation values of other physical observables to confirm that the prepared state captures well the properties of the exact ground state. As an example, we calculated the spin-spin correlation in $x$-direction $\langle \sigma^x_i \sigma^x_j \rangle$ for the 12-qubit lattice without diagonal gates and for open boundary conditions. We compare the result with the spin-spin correlation function of the exact result and calculate the difference, as can be seen in figure \ref{fig: Correlation_without_NN}.
We can see a very good agreement, showing that the VQE approximation is indeed very accurate.

\subsection{Variation of $J_2$
\label{sec:variationj2}}
\begin{figure}
    \centering
    \includegraphics[width=0.4\textwidth]{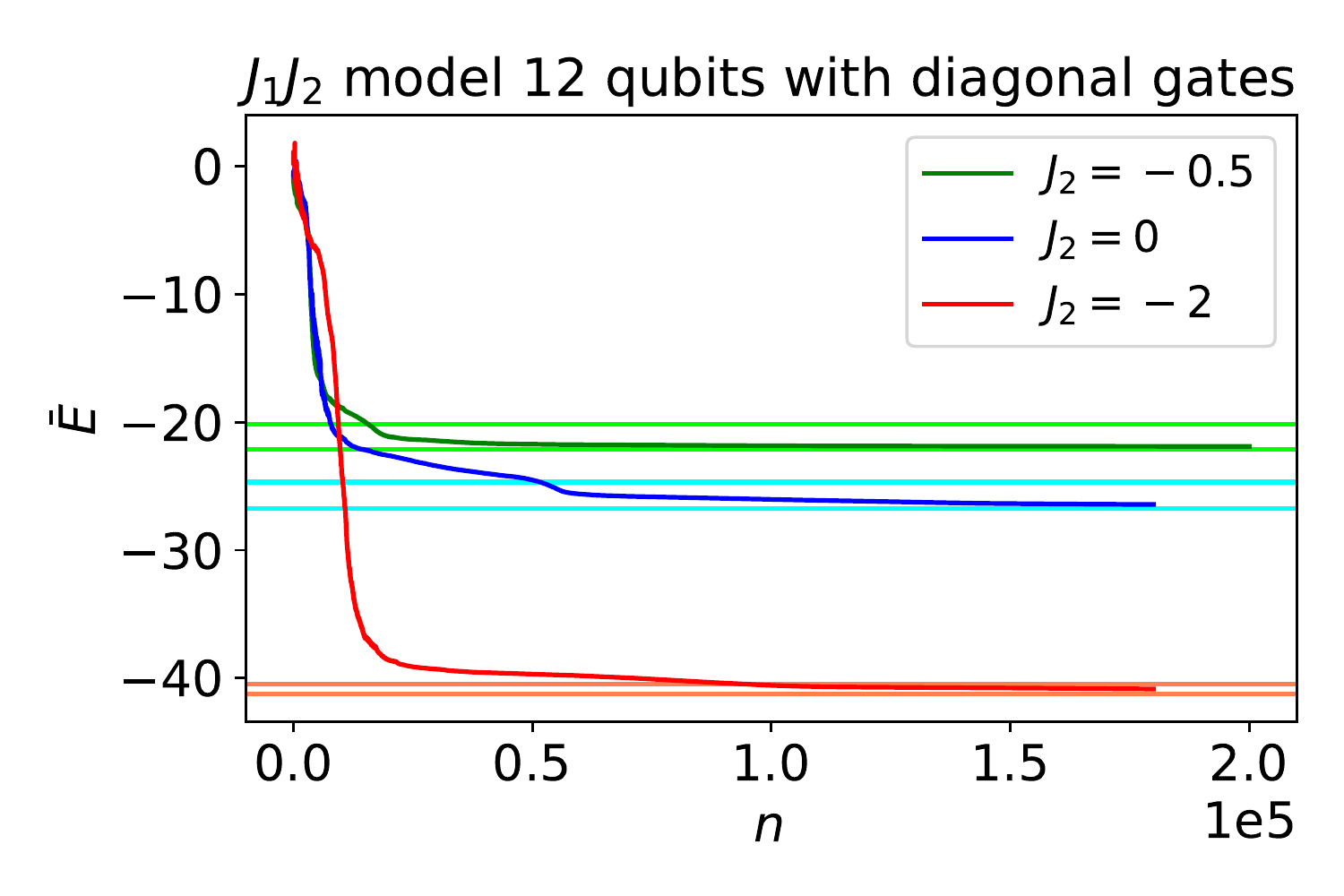}
    \caption{VQE performance for the $J_1-J_2$-model with 12 qubits without diagonal interaction gates. The lighter lines mark the ground state energy $E_0$ and energy of the first excited state $E_1$.
    The circuit for each configuration of $J_2$ consists of 7 gate layers plus the additional X- and Y-layer at the beginning.}
    \label{fig:j2var}
\end{figure}

To explore how good the ansatz performs for different values of the ratio of the couplings $J_2/J_1$, and therefore for different quantum phases of the model, we here show results for a fixed value of $J_1=-1$ where we varied the value of $J_2$, see figure \ref{fig:j2var}. 
We choose the range of $J_2$ values such that the different quantum phases of the $J_1-J_2$-model are covered.
For the collinear Néel ordered states we choose $J_2 = -2$ and for the Néel ordered ground state $J_2 =0$, which thus corresponds to the Heisenberg model. The obtained values for the energies $\overline{E}$ can be found in table \ref{tab:results}.

For all choices of $J_2$, our ansatz achieves a good convergence to the respective ground state. For the ground state energy for the VQE with $J_2 =0$, we achieve a value of $16 \%$ of the spectral gap. For $J_2= -2$ the VQE achieved a value of only $48\%$ of the spectral gap. We note that a good convergence in terms of the spectral gap is in this case difficult to achieve because the spectral gap is very small. Nonetheless, the achieved energy is close to the ground state energy. Our findings are thus in agreement with the generic behavior that in cases where the spectral gap is narrow, more gate layers have to be used to achieve a higher accuracy. 
We can thus see that the ansatz is rather versatile and yields good approximations for all phases of the model. We attribute this good performance to the choice of the ansatz, including the parametrized X-and Y-gates at the beginning of the circuit.

\subsection{Extrapolation of parameter-numbers for larger lattice sizes \label{sec:extrapolate}}

The effort of running a VQE algorithm is determined by the number of gates that are needed, since this number determines the hardware requirements, and the number of variational parameters that are needed, since this number determines the number of optimization steps and thus the number of required measurements.

To estimate how many gates and how many variational parameters are needed in our ansatz for larger systems, we plotted the number of required gate numbers and variational parameters for the system sizes for which we did our simulations and extrapolated the resulting curve for each ansatz. As a criterion for successful convergence, we require that the achieved expectation value $\overline{E}$ is lower or equal to the middle of the spectral gap $E_1 - E_0$. We then determine the minimal number of gate layers or minimal number of gates needed to achieve this, and plot the corresponding number of two-qubit gates and variational parameters numbers for all lattice sizes for which we can simulate the algorithm classically. By fitting a polynomial to this data, we extrapolate the expected required number of two-qubit gates and variational parameters to larger lattice sizes.

\begin{figure}
    \centering
    \includegraphics[width=0.4\textwidth]{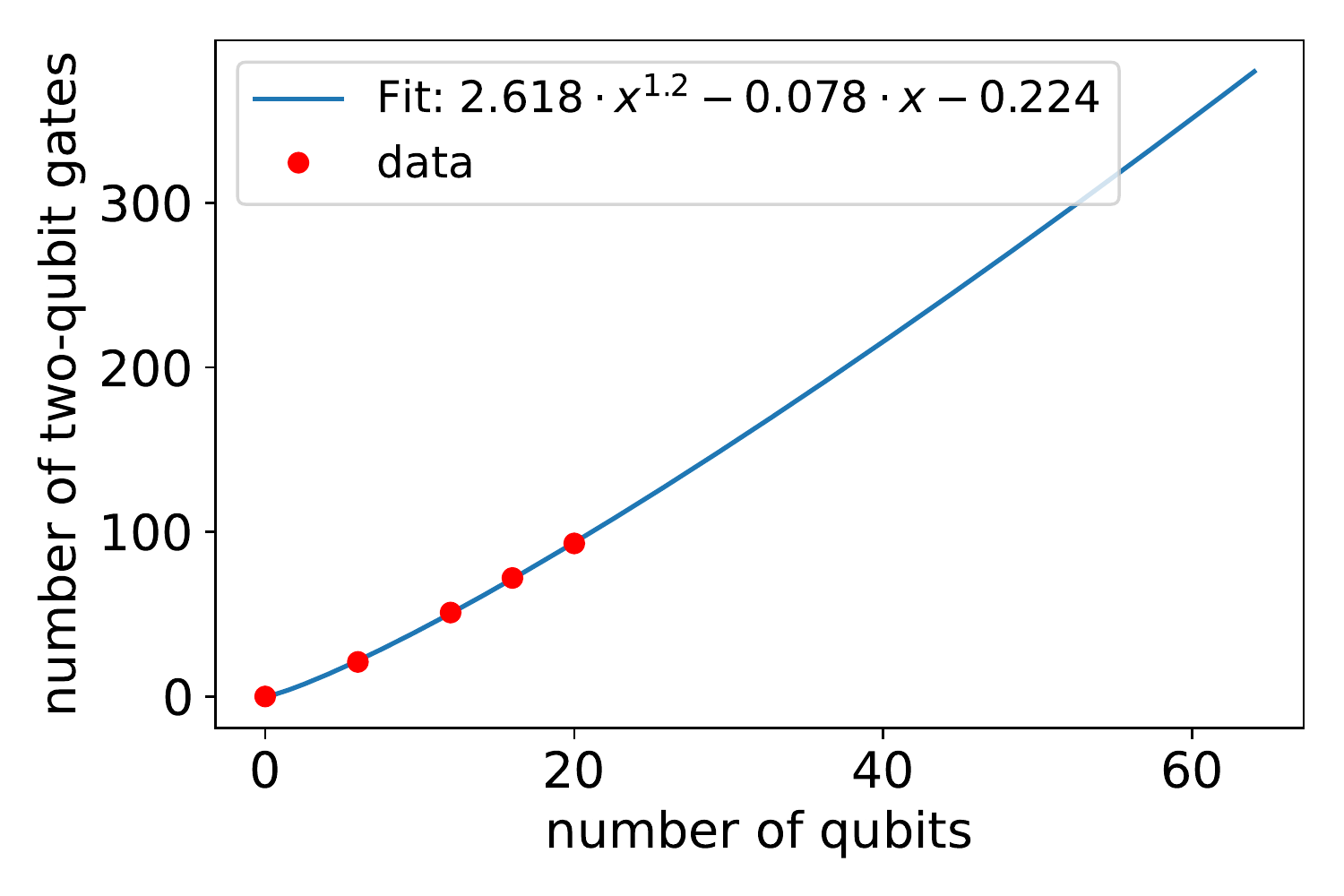}
    \caption{Number of two-qubit gates required in our numerical studies using the ansatz without diagonal-interaction gates to achieve $\overline{E}\leq 0.5\cdot E_1 - E_0$ (red dots) for different lattice sizes and fit to this data (blue line). The extrapolation of the fit shows the expected number of two-qubit gates for up to 64 qubits (corresponding to an $8 \times 8$ qubit grid). We find a $x^{1.2}$-growth of the required number of parameters. The number of two-qubit gates here is the sum of all XX-, YY- and ZZ-gates used to achieve the required convergence. 
    \label{fig:extrapolate_2qubitgates}}
\end{figure}

\begin{figure}
    \centering
    \includegraphics[width=0.4\textwidth]{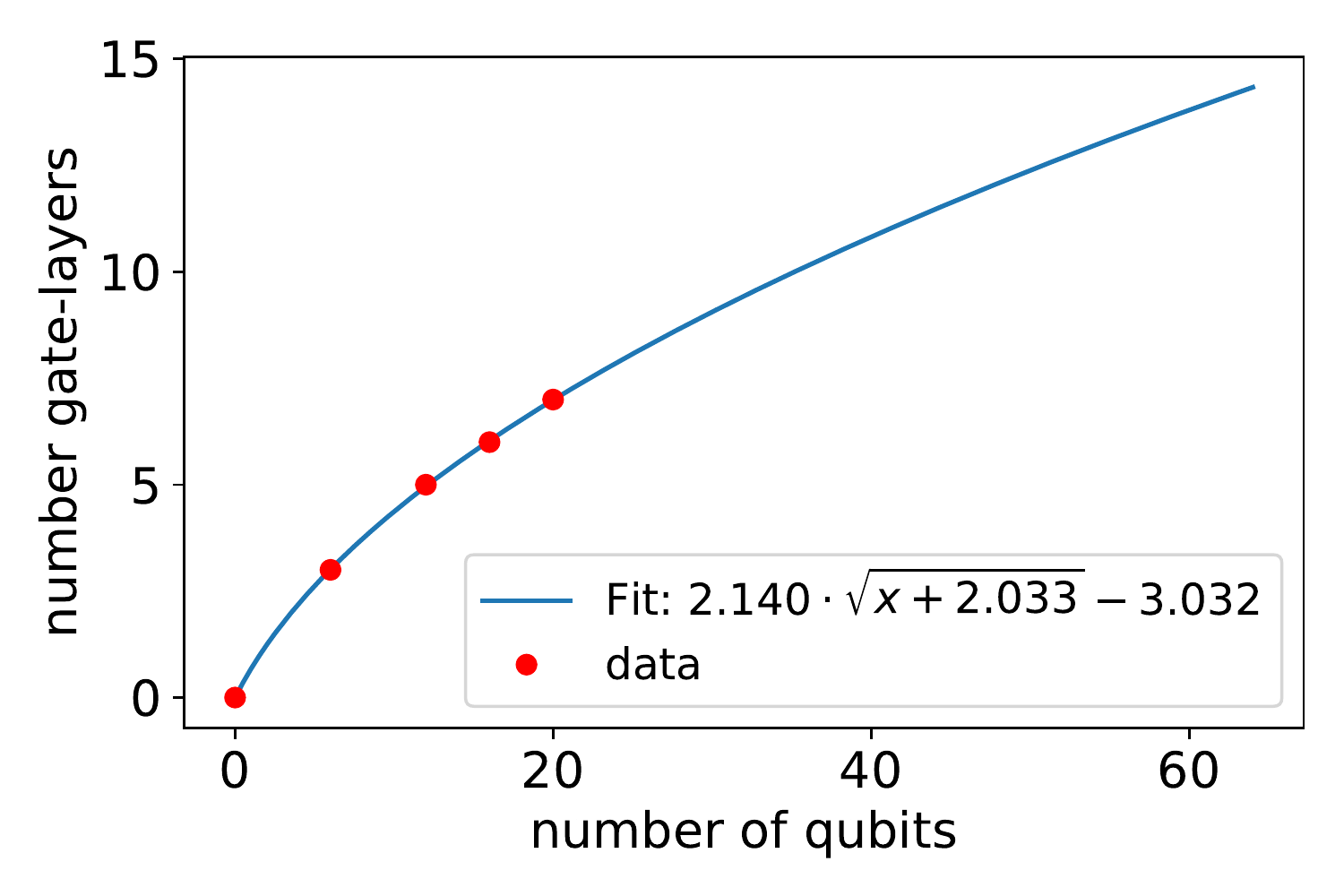}
    \caption{Number of gate layers required in our numerical studies to achieve $\overline{E}\leq 0.5\cdot E_1 - E_0$ (red dots) for different lattice sizes and fit to this data (blue line). The extrapolation of the fit shows the expected number of gate layers for up to 64 qubits (corresponding to an $8 \times 8$ qubit grid). We find a $\sqrt{x}$-growth of the required number of parameters. This plot is valid for both ans\"atze (with and without diagonal gates).}  
    \label{fig:extrapolate_layers}
\end{figure}

The number of required gates to achieve $\overline{E}\leq 0.5\cdot E_1 - E_0$ for the efficient ansatz without diagonal gates is shown in figure \ref{fig:extrapolate_2qubitgates}. We find a $n^{1.2}$ dependency on the qubit number. These gate numbers correspond to the required number of gate layers as shown in figure \ref{fig:extrapolate_layers}, which is identical of both ans\"atze (with and without diagonal gates) due to the definition of gate layer that we use. Interestingly, the number of required layers grows less than linear in the number of qubits, which is a promising behavior for scalability.

\begin{figure}
    \centering
    \includegraphics[width=0.41\textwidth]{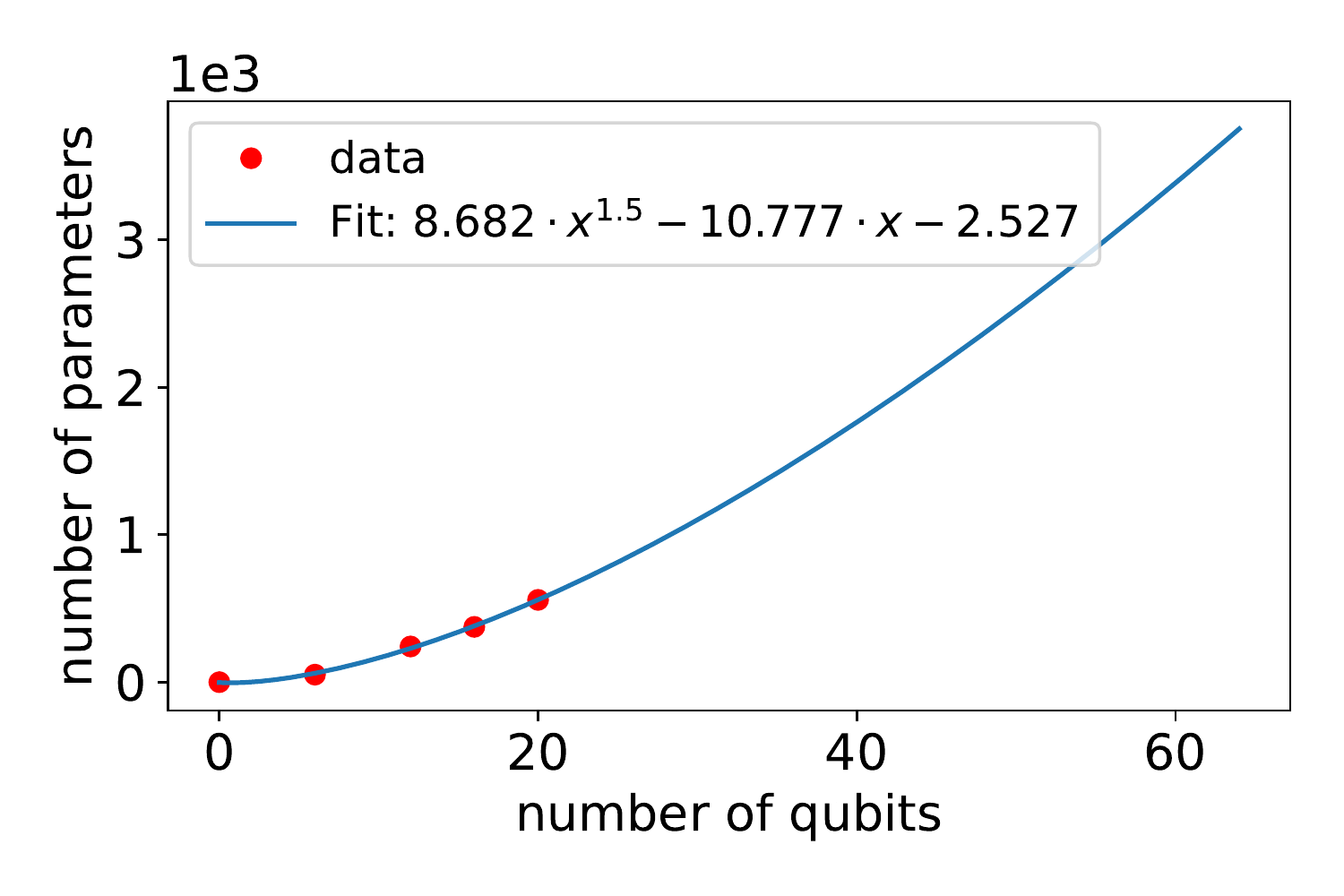}
    \caption{Number of variational parameters used in our numerical studies using the ansatz with diagonal-interaction gates for different lattice sizes (red dots) and fit to this data (blue line). The extrapolation of the fit shows the expected number of parameters for up to 64 qubits (corresponding to an $8 \times 8$ qubit grid). We find an $x^{1.5}$-growth of the required number of parameters.}  
    \label{fig:extrapolate_with}
\end{figure}

\begin{figure}
    \centering
    \includegraphics[width=0.41\textwidth]{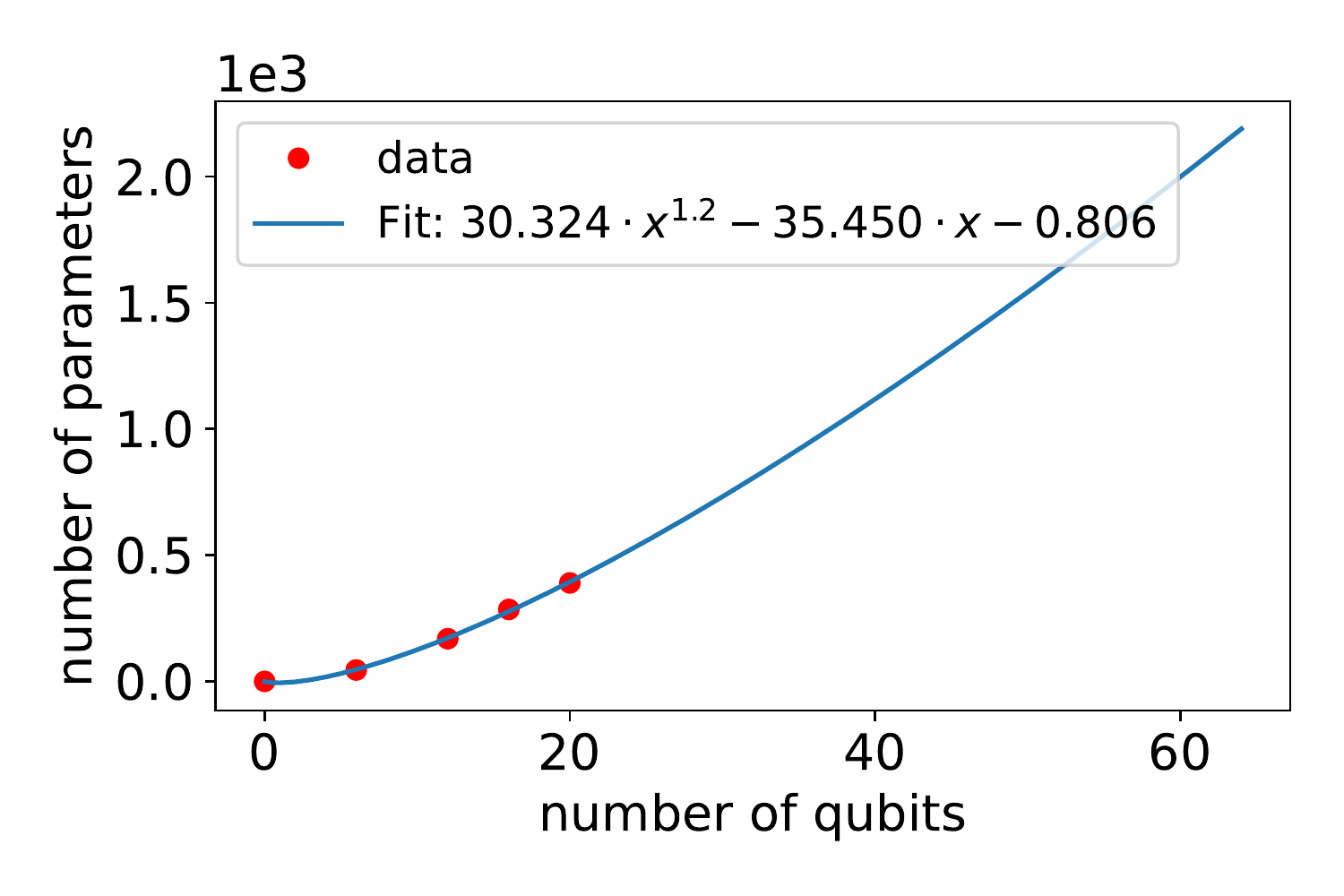}
    \caption{Number of variational parameters used in our numerical studies using the ansatz without diagonal-interaction gates for different lattice sizes (red dots) and fit to this data (blue line). The extrapolation of the fit shows the expected number of parameters for up to 64 qubits (corresponding to an $8 \times 8$ qubit grid). Using this extrapolation, we find an $x^{1.2}$-growth of the required number of parameters.}
    \label{fig:extrapolate_without}
\end{figure}

The number of required variational parameters to achieve $\overline{E}\leq 0.5\cdot E_1 - E_0$ is shown in figure \ref{fig:extrapolate_with} for the ansatz with next-nearest neighbor gates and in figure \ref{fig:extrapolate_without} for the ansatz without next-nearest neighbor gates. With our extrapolation procedure, we find a $n^{1.5}$ dependency on the qubit number $n$ for the number of parameters of the ansatz with diagonal interaction gates and a dependency of $n^{1.2}$ for the ansatz without the diagonal interaction gates. Hence, the curve for the ansatz without the  diagonal gates increases more slowly than for the ansatz with diagonal gates. This observation indicates that omitting the diagonal gates can lead to a significant reduction in the depth of the ansatz and thus the circuit, as one increases the size of the spin lattice.


\section{Conclusions}
We simulated and tested a Variational Quantum Eigensolver for the two-dimensional $J_1-J_2$-model. Using the proposed ansatz, one can access the ground state energy of this spin-model. Moreover, our ansatz can be used without gates that directly implement the diagonal next-nearest neighbor interactions. This feature allows to avoid SWAP gates when executing the gate sequence on hardware with qubits on a rectangular grid that can only undergo gate operations with their nearest neighbors.
Moreover, we analyzed the different quantum phases in the $J_1-J_2$-model model by varying $J_2/J_1$.
We saw that our ansatz worked well for all the different configurations and led to a sufficient 
accuracy in the ground state approximation. 
To improve the performance, a better suited optimizer than COBYLA or more basinhopping iterations could be used to avoid getting stuck in energy plateaus or local minima. Another option to improve the performance, are greater circuit depths with more parameters or a more suitable initial configuration that already displays some information about the system. 

By fitting a polynomial to the required number of two-qubit gates and the required number of variational parameters versus qubit number, we anticipate that the required number of two-qubit gates and parameters grow slower than quadratically in the number of qubits. While the required number of gates can not yet be run with sufficient accuracy on existing hardware, successful VQE implementations that eventually may challenge results of classical numerics thus seem within reach in the near future.

\begin{acknowledgments}
This work received support from the German Federal Ministry of Education and Research via the funding program quantum technologies - from basic research to the market under contract number 13N15684 ``GeQCoS" and under contract number 13N15577 ``MANIQU". It is also part of the Munich Quantum Valley, which is supported by the Bavarian state government with funds from the Hightech Agenda Bayern Plus.
\end{acknowledgments}

\appendix

\section{Details of the Simulation}
The simulation of the Variational Quantum Eigensolver for the $J_1-J_2$-model was done in Python 3.8.5 with the help of NumPy 1.20.2 \cite{numpy} using Cirq 0.10.0 \cite{cirq_developers}.  For the optimization, the built-in optimizers from SciPy 1.6.3 \cite{scipy} were used.
We used the gate \texttt{cirq.X(q), cirq.Y(q), cirq.Z(q) cirq.XX(q1,q2), cirq.YY(q1,q2)} and \texttt{cirq.ZZ(q1,q2)} built in Cirq to a power of the respective parameter. And defined the Hamiltonian via Pauli operators in Cirq via \texttt{cirq.PauliSum}.
After applying a sufficient amount of gate-repetitions, the circuit is simulated. 
We used qsimcirq, a full wave function simulator written in C++ which is much faster than the normal simulator in Cirq.

\section{Results for periodic boundary conditions}
For the 12-qubit model we also tested if the ansatz works for a lattice with periodic boundary conditions (In figure \ref{fig:16grid} periodic boundaries would mean for example a $J_1$ interaction between qubits 1 and 4, 1 and 13. As well as $J_2$ interactions, e.g. of qubits 1 and 14 and qubits 2 and 13).

For this lattice with periodic boundary conditions, the exact values for the ground and first excited state energies can be found in table \ref{tab:results}. We achieve a value for $\bar{E}$ for which $\frac{\bar{E}-E_0}{E_1-E_0} <10 \% $. We thus conclude that our ansatz also works with periodic boundary conditions and the same number of gate layers (see table \ref{tab:params}) as for the lattice with open boundaries.

\begin{figure}
    \centering
    \includegraphics[width=0.4\textwidth]{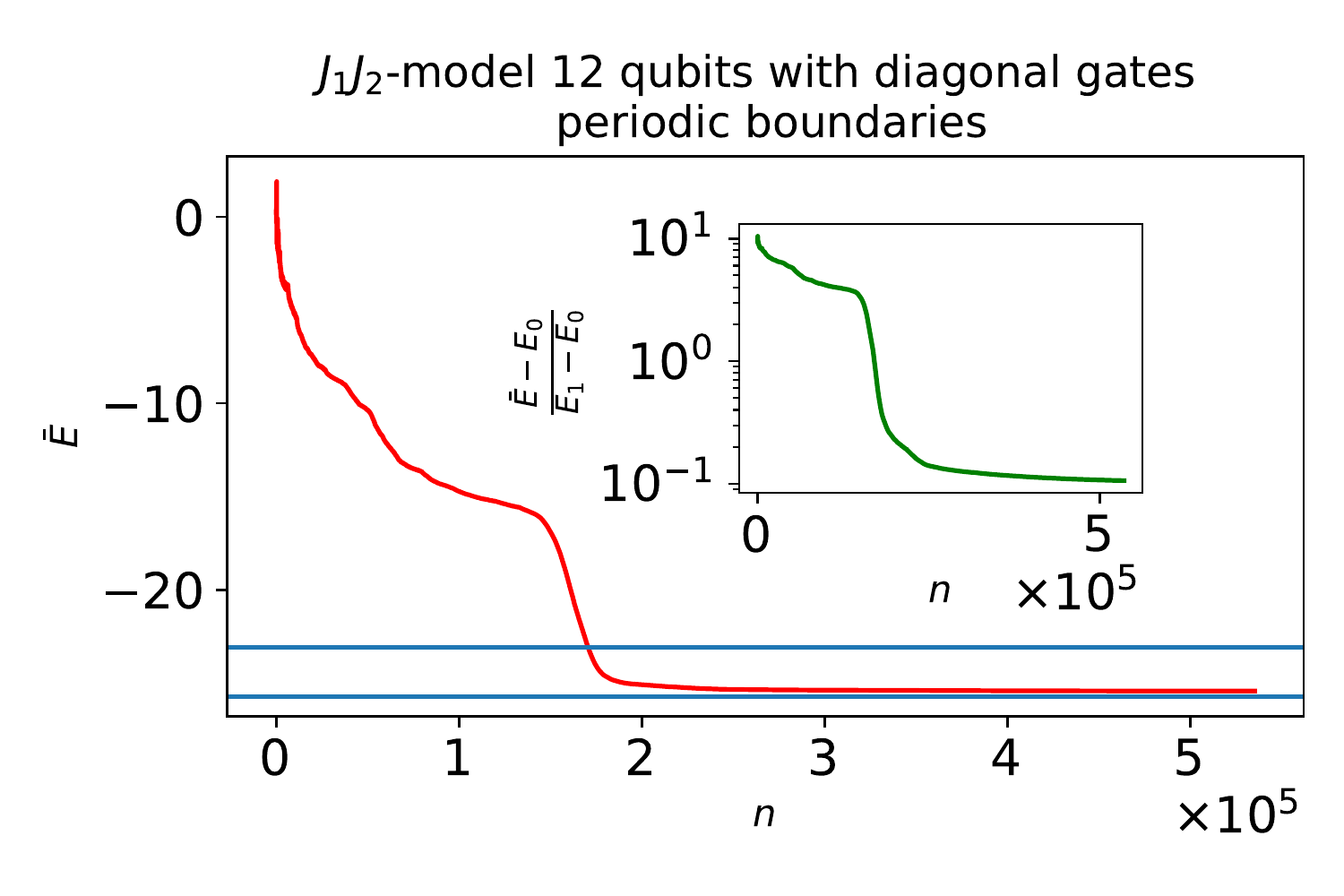}
    \caption{VQE performance for the $J_1-J_2$-model with 12 qubits with periodic boundary conditions. The lower blue line marks the ground state energy $E_0$ and the upper line the energy of the first excited state $E_1$. The circuit consists of 7 layers of gates plus the additional X- and Y-layer at the beginning.
    The inset shows the run of the VQE-steps against the difference of the expectation value $\bar{E}$ from the VQE and the exact ground state energy $E_0$ divided by the spectral gap, which is the difference of the energy of the first excited state $E_1$ and the ground state energy.}
    \label{fig:12periodbound}
\end{figure}

\nocite{*}

\bibliography{mybib}

\end{document}